\documentclass[12pt]{article}

\usepackage{amsfonts,amssymb,latexsym,amsthm,amscd}

\usepackage{amsmath,amsthm,amssymb}
\usepackage{graphics,graphicx}
\usepackage{epsfig}
\newfont{\tenmsb}{msbm10 scaled\magstep1}

\newcommand{\half}{{\scriptstyle{\frac{1}{2}}}}

\newcommand{\cC}{{\cal C }}

\newcommand{\parag}{\hfil\break}
\newcommand{\kikezd}{\parag\underbar}

\def\smallover#1/#2{\hbox{$\textstyle{#1\over#2}$}}
\def\p{{\partial}}
\def\lf{\left(}
\def\rg{\right)}
\def\lq{\left[}
\def\rq{\right]}
\def\lgr{\left\{}
\def\rgr{\right\}}
\def\va{{\vec a}}
\def\vd{{\vec d}}

\def\vb{{\vec b}}

\def\vE{{\vec E}}
\def\vB{{\vec B}}

\def\vk{{\vec k}}

\def\vp{{\vec p}}

\def\vq{{\vec q}}
\def\vr{{\vec r}}

\def\vA{{\vec A}}
\def\vB{{\vec B}}

\def\cR{{\cal R}}
\def\cQ{{\cal Q}}
\def\cT{{\cal T}}
\def\cE{{\cal E}}
\def\cA{{\cal Q}}
\def\vpi{{\vec \pi }}
\def\he{{\hat{\epsilon}}}

\def\cP{{\cal P}}

\def\cJ{{\cal J}}

\def\cK{{\cal K}}
\def\cE{{\cal E }}

\def\*{{\star}}
\def\beq{\begin{equation}}
\def\eeq{\end{equation}}
\def\beqa{\begin{eqnarray}}
\def\eeqa{\end{eqnarray}}
\def\nn{\nonumber}
\def\Sch{Schr\"odinger equation  }
\begin{document}

\setlength{\baselineskip}{16pt}

\title{Dynamics of semiclassical Bloch wave - packets }

\author{P.~A.~Horv\'athy\footnote{e-mail: horvathy@lmpt.univ-tours.fr}
\\
Laboratoire de Math\'ematiques et de Physique Th\'eorique\\
Universit\'e de Tours\\
Parc de Grandmont, F-37 200 TOURS (France)
\\
L. Martina\footnote{e-mail: Luigi.Martina@le.infn.it}
\\
Dipartimento di Fisica dell'Universit\`a and Sezione INFN di
Lecce. \\ Via Arnesano, CP. 193 I-73 100 LECCE (Italy). }

\date{}

\maketitle

\begin{abstract} The semiclassical approximation for electron
wave-packets in crystals leads to equations which can be derived
from a  Lagrangian or, under suitable regularity conditions,
in a Hamiltonian framework.
In the plane, these issues are studied %in presence of external fields
 using the method of the coadjoint orbit
applied to the ``enlarged'' Galilei group.
%The relation with ``exotic'' models, associated with the
%2-parameter central extension of the Galilei group, is indicated.
\end{abstract}

%%%%%%%%%%%%%%%%%%%%%%
\section{Introduction}
%%%%%%%%%%%%%%%%%%%%%%
The standard semiclassical dynamics of a Bloch electron in a solid
\cite{Ashcroft}  accounts for various properties of metals,
semiconductor and insulators.  More recently it was argued,
however,
 that  a correct wave packet dynamics
requires  taking into account
also geometric (Berry) phase effects \cite{Niu}. The latter modify  the transport
properties of metals and semiconductors, and provide us with a new insight into
the Anomalous Hall Effect  \cite{AHE,Fang}.

From the theoretical point of view, two
problems arose~: i) the accuracy of the semiclassical approximation
 derived from a time dependent
variational principle in Quantum  Mechanics \cite{Kramer}; ii) the
geometrical  structure of the dynamical systems describing the
evolution of the electron wave-packets. The %natural
Hamiltonian structure of these models %, may indeed
% be hidden by their non-canonical symplectic structure.
%This  has a deep physical meaning,
%since the non-commutativity of the position coordinates is related
takes into account  the Berry Phase effects by non commuting
coordinates and realizes, at least in the planar case,  a two-fold
central extension %a nontrivial extension
 of the Galilean symmetry \cite{exotic}.

% The planar Galilei group admits a two-fold central extension.
 %labeled by the mass $m$  and a second, ``exotic'',  parameter $\kappa$
%\cite{exotic}.
For uniform electromagnetic fields, the structure can further be
extended~: position-independent fields can be viewed as extra
``coordinates'' that can be added to the ordinary space-time
variables. The symmetries of the combined structure form the
``enlarged'' or ``Maxwell-Galilei'' group \cite{Negro,HMS}, which
involves, besides the usual Galilean space-time symmetries, also
field components, see (\ref{enlargedGal}).

Firstly, physical realizations of such a symmetry have been
presented in \cite{LSZ,DH}.  Particles of this type  are related
to ``anyons'', and may be used in explaining the Fractional
Quantum Hall Effect \cite{Laughlin}. Difficulties arise when
coupling to an external electromagnetic field, but this can be
partially overcome
  by resorting to the methods  %relativistic and non-relativistic
 of the coadjoint
orbits on  a larger symmetry group \cite{Negro}.
%The framework applies to both free particles
%and also to particles interacting with a uniform electromagnetic field.
%On the other hand,  while anyons are related to 2 space dimensions,
% concept of does not seem very useful
Similar symplectic structures may appear also in 3 space
dimensions \cite{BM,HMS2}.

In Sec. 2  we briefly  review the main ideas involved in the
semiclassical approximation of Bloch electron wave-packets.
%leading to dynamical models of interest.
 In Sec. 3  their
Lagrangian and Hamiltonian formulations are considered. Then
in Sec. 4,
considering a simplified version (in the plane but interacting
with constant external fields),  we study their general
 geometric formulation, by resorting to the coadjoint orbit
method \cite{SSD}.
%Final considerations are presented in the Discussion
%Section 5.

%%%%%%%%%%%%%%%%%%%%%%%%%%%%%%%%%%%%%%%%%%
\section{Semiclassical Approximation}
%%%%%%%%%%%%%%%%%%%%%%%%%%%%%%%%%%%%%%%%%%
The \Sch equation can be derived from  the action
functional \beq S=\int_{t_1}^{t_2} L_S dt,
%, \eeq with Lagrangian
\qquad
% \beq
L_S = \frac{i}{2}
 \frac{\langle \Psi| \frac{d \Psi}{d t}\rangle -
  \langle \frac{d \Psi}{d t}| \Psi \rangle }
{\langle \Psi| \Psi \rangle } - \frac{\langle \Psi|{\hat H}| \Psi
\rangle }{\langle \Psi| \Psi \rangle}  ,\label{SchLagr}\eeq
requiring the action to be  stationary at the ``classical" wave function
 $| \Psi\rangle$. The latter belongs to  a suitable
 Hilbert  space, acted uponby an  hermitian Hamiltonian
  operator ${\hat H}$. The   usual \Sch  is obtained  after a
 suitable phase normalization \cite{Kramer}.

 The  derivation of the semiclassical approximation from a variational principle requires
 restricting ourselves  to a predetermined domain of the
 Hilbert space by a suitable  parametrization
 of the wave-function $|\Psi\rangle$, such  that the variational
 principle singles out the best approximate time evolution.
 In particular, for a spinless point-like particle one can introduce
 the mean position and momentum values,
 $$ \vr_c \lf t \rg =  \frac{\langle \Psi|{\hat \vr}| \Psi
\rangle }{\langle \Psi| \Psi \rangle},
\qquad
 \vp_c \lf t \rg =
\frac{\langle \Psi|{\hat \vp}| \Psi \rangle }{\langle \Psi| \Psi
\rangle}
$$ as
 main parameters. Neglecting all other details of the
time evolution of the wave-packets,   we  replace $|\Psi\rangle
\to |\widetilde{\Psi}\lq \vr_c\lf t \rg, \vp_c\lf t \rg\rq
\rangle$ into the  Lagrangian (\ref{SchLagr}) and  look for
self-consistent equations for the parameters.
%By enlarging
%the number of variational parameters, one  may  brought closer
%$|\widetilde{\Psi}\rangle $ to the generic $|\Psi\rangle$. But,
%since this parametrized subdomain of the Hilbert space does not
%have the structure of a linear subspace,  the superposition of
%states is meaningless. Now the general ideology of such an
%approach consists in assuming that the evolution of the considered
%states is restricted by a (stability) subgroup of a symmetry Lie
%group. This implies that all expectation values of the observables
%of the the system actually take values in a coset space
%determined by the group structure, and the variational parameters
%naturally evolve in a symplectic manifold \cite{Kramer}.
%All these arguments work very well for compact Lie groups, which
%is not the case for the Bloch electrons system, though. Lacking a
%deeper mathematical analysis,
 %The procedure is justified \emph{a
%posteriori} by the physically consistent results obtained for the quantum
%Hall effect and for the motion of magnetic Bloch states \cite{Niu}.

As usual, the dynamics of an electron in a crystal lattice is
described by a first quantized Hamiltonian operator \beq  {\hat
H}\lq {\hat\vr}, {\hat \vp}, f \lf {\hat\vr}, t \rg \rq,
\label{HamOr}\eeq
 where
$\hat\vr$ and $ {\hat \vp}$ are  position and momentum operators,
 satisfying the Heisenberg algebra. The (possibly vectorial and/or time dependent)
function $f$ represents  differentiable ``slow" modulations in
space with respect to a fixed lattice background potential, such
that %\beq
${\hat H}_c = {\hat H}\lq {\hat\vr}, {\hat \vp}, c \rq$
%\label{perfHam}\eeq
is a periodic Hamiltonian operator  for any constant parameter
$c$. To justify the previous parametrization,  three
phenomenological length scales have to be considered, namely the
typical lattice constant length $l_{latt}$, the wave-packet
dispersion length $l_{w p}$ and the modulation wave-length
$l_{mod}$. They are related by  $l_{latt} \ll l_{w p} \ll
l_{mod}$. Assuming regular dependence of the Hamiltonian operator
(\ref{HamOr}) on the modulations, one can truncate it at the first
order around the instantaneous mean position $\vr_c$, yielding
\beq {\hat H} = {\hat H}_{\lf \vr_c, t \rg} + \half \lq \p_f {\hat
H}\, \nabla_{\vr_c } f \lf {\vr_c}, t \rg \cdot \lf {\hat \vr} -
\vr_c \rg + h.c. \rq ,\label{HamApprox}\eeq where ${\hat H}_{\lf
\vr_c,t \rg}$ belongs to the  family  of the ${\hat H}_c$, with $c
= \lf {\vr_c}, t \rg$. Thus, we postulate that the physical states
does not only include a (approximate) solution  of the \Sch for a
special value of $\vr_c$, but for all possible $\vr_c\lf t \rg $
belonging to some configuration space, which has to be determined
self-consistently. Then, for any $\vr_c$, one can choose an
orthonormal basis of eigenvectors of the Bloch parameter-dependent
Hamiltonians  ${\hat H}_{\lf \vr_c,t \rg}$, that is \beq {\hat
H}_{\lf \vr_c,t \rg} |\psi_{\lf \vr_c,t \rg}^{n ,\vq}\rangle =
E_{\lf \vr_c,t \rg}^{n ,\vq} |\psi_{\lf \vr_c,t \rg}^{n
,\vq}\rangle,\quad \langle \psi_{\lf \vr_c,t \rg}^{n ,\vq}
|\psi_{\lf \vr_c,t \rg}^{n' ,\vq \, '}\rangle = \delta_{n,n'}
\delta \lf \vq - \vq\,'\rg, \label{eigenprob}\eeq
 where the energy eigenvalues $E_{\lf \vr_c,t \rg}^{n ,\vq}$
are labelled by a band index $n$ and by the quasi-momentum $\vq$,
which can be limited to belong to the first Brillouin zone (IBZ).
In the position representation the Bloch eigenfunctions take the
form \beq \langle \vr|\psi_{\lf \vr_c,t \rg}^{n ,\vq}\rangle =
e^{i \vq \cdot \vr } u_{\lf \vr_c,t \rg}^{n ,\vq} \lf \vr \rg  ,
\qquad u_{\lf \vr_c,t \rg}^{n ,\vq} \lf \vr + {\vec a}\rg = u_{\lf
\vr_c,t \rg}^{n ,\vq}\lf \vr \rg , \label{Bloch_wf}\eeq $u_{\lf
\vr_c,t \rg}^{n ,\vq} \lf\vr \rg$ being the periodic part of the
wave-function, assumed to be
 analytic in $\lf \vr_c,t \rg$. In the
eigenvalue problem (\ref{eigenprob}) time dependence is assumed
adiabatic, so that the eigenvalues $E_{\lf \vr_c,t \rg}^{n ,\vq}$
form  well separated bands. This is not always the case, but one
can assume that the degeneracy occurs in isolated points of the
IBZ not considered in the present discussion. However, there are
effects which could depend on the detailed behavior of
(quasi)-degenerate bands, if the Fermi level is close to these
values \cite{Fang}.  Moreover, we assume that the  modulations are
so weak that band jumping is forbidden.
%This can be expressed by requiring that the typical
%modulation time scale is $t^n_{mod} \gg \frac{\hbar}{E_{gap,n}}$,
%where $E_{gap,n}$ is the n-th energy gap. Analogously, one can
%require that the space modulation scale be bounded from below by
%$l_{mod} \gg \frac{\hbar}{\texttt{Im}\lf \vq_{gap, \, n}\rg}$.

Now, it is a classical result by Karplus and Luttinger \cite{Niu}
that  the matrix element of the position operator between two
Bloch wave-functions is given by \beq \langle \psi_{\lf \vr_c,t
\rg}^{n ,\vq}|\,{\hat \vr}\,| \psi_{\lf \vr_c,t \rg}^{n' ,\vq'}
\rangle = \lq i \nabla_{\vq}  +  \, \langle u_{\lf \vr_c,t \rg}^{n
,\vq} \lf \vr \rg | i \nabla_{\vq} \,u_{\lf \vr_c,t \rg}^{n ,\vq
\,} \lf \vr \rg\rangle_{cell} \rq \, \delta \lf \vq\,' -
\vq\rg\,\delta_{n, n'}. \label{MatrixEl}\eeq That is, in the space
generated by the Bloch waves,  ${\hat \vr}$ acts as \beq {\hat
\vr} = i \nabla_{\vq} + {\vec{\cA}}\,^{n }{\lf \vr_c, \vq ,t \rg},
\quad {\rm with }\quad{\vec{\cA}}\,^{n }{\lf \vr_c, \vq ,t \rg} =
\langle u_{\lf \vr_c,t \rg}^{n ,\vq} | i \nabla_{\vq} \, u_{\lf
\vr_c,t \rg}^{n ,\vq} \rangle_{cell},
 \label{BerryConn} \eeq
where
$\langle \cdot|\cdot \rangle_{cell}$ is the restriction of the
scalar product for ${\cal L}^2_{cell}$  to the unite
cell with periodic boundary conditions.  $\lf 2 \pi \rg^3/
V_{cell}$ is a normalization factor.
The quantity $\vec{\cA}\,{}^{n}{\lf \vr_c, \vq ,t \rg}$
here  can be interpreted as a $U\lf 1 \rg $ gauge connection and is
identified with \textit{Berry's connection}.

The components of the position operator no longer commute in
general, \beq \lq {\hat r}_j, {\hat r}_l \rq =  i \lf \p_{q_j}
{{\cA}}\,^{n }_l {\lf \vr_c, \vq ,t \rg} - \p_{q_l} {{\cA}}\,^{n
}_j {\lf \vr_c, \vq ,t \rg}\rg \equiv i \; \Theta^{n }_{j, l}\lf
\vr_c, \vq, t \rg, \eeq where the antisymmetric tensor \beq
\Theta^n\lf \vr_c, \vq, t \rg = \lf \Theta^{n }_{j, l} \rg =
 i \langle \nabla_{\vq} u_{\lf
\vr_c,t \rg}^{n ,\vq} | \times |  \nabla_{\vq} \, u_{\lf \vr_c,t
\rg}^{n ,\vq} \rangle_{cell}  \label{BerryTens}\eeq
is the gauge invariant \textit{Berry curvature}.

Thus,  the Berry phase converts the dynamics  of an ordinary
particle, in a periodic background potential,  into  a quantum
mechanical system living in a  non-commutative configuration space
 \cite{Szabo}.
%We  do not discuss further this aspect, limiting ourselves to remarking that, following a closed path in
%quasi-momentum space while holding   $\vr_c$ fixed
%in the configuration space makes its eigenfunction to acquire a
%net phase shift determined by the Berry curvature.

Now we want to build  wave-packets from the Bloch wave-functions
(\ref{eigenprob} - \ref{Bloch_wf}) chosen from a single energy
band (say $n$; we drop the index in what follows)  of the form
\beq |\widetilde{\Psi}\lq \vr_c\lf t \rg, \vq_c\lf t \rg\rq
\rangle = \int_{_{IBZ}}  \Phi\lf \vq, t \rg | \psi_{\lf \vr_c,t
\rg}^{\vq } \rangle \; d\, \vq , \label{AppWF}\eeq where the
quasi-momentum normalized amplitude $\Phi\lf \vq, t \rg $ ( taking
$\int_{_{IBZ}} | \Phi\lf \vq, t \rg |^2 \; d\, \vq = 1$) is chosen
in such a way that its dispersion $\Delta_q$ in momentum space is
small compared to the first typical Brillouin  size $\sim 2 \pi /
l_{latt}$. Moreover, we  describe  the semiclassical approximate
wave function by the mean quasi-momentum \beq \vq_c \lf t \rg =
\int_{_{IBZ}} \vq \; | \Phi\lf \vq, t \rg |^2 \; d\, \vq , \eeq
completed with the mean position \beqa \vr_c \lf t \rg =
  \langle \widetilde{\Psi}\lq \vr_c\lf t \rg, \vq_c \lf t \rg\rq | \,
  {\hat {\vr}}\, |
  \widetilde{\Psi}\lq \vr_c\lf t \rg, \vq_c\lf t \rg\rq
  \rangle \nn \\
  = \int_{_{IBZ}} |\Phi\lf \vq, t \rg|^2 \, \lq - \nabla_{\vq}\,
  {\rm arg} \lq \Phi\lf \vq, t \rg \rq  +
  {\vec{\cA}}\,{\lf \vr_c, \vq ,t \rg }\rq d \, \vq
  \\ \approx  - \nabla_{\vq_c}\,
  {\rm arg} \lq \Phi\lf \vq_c, t \rg \rq  +
  {\vec{\cA}}\,{\lf \vr_c, \vq_c ,t \rg } \label{ApproxPos}
 \eeqa
were the relation (\ref{MatrixEl}) has been used. In the last
  approximate equality  only contributions of zero
 order in the wave-packet space and momentum dispersion lengths,
 $l_{w p}$ and $\Delta_q$ respectively,
 were retained. This is the meaning of the
 semiclassical approximation, which we use to
 evaluate  the  Lagrangian
 (\ref{SchLagr}) after inserting the wave-function
 (\ref{AppWF}). Thus,
 one finds the following approximate relations:
 \beqa
  \langle \widetilde{\Psi} | \, i \frac{d}{d t}\, | \widetilde{\Psi}\rangle
    & \approx - \p_t   {\rm arg} \lq \Phi\lf \vq_c, t \rg \rq  +  \langle u_{\lf
\vr_c,t \rg}^{\vq_c} | i \p_t \, u_{\lf \vr_c,t \rg}^{\vq_c}
\rangle_{cell}  + \dot{{\vr_c}}  \cdot \langle u_{\lf \vr_c,t
\rg}^{\vq_c} |i  \nabla_{\vr_c} \, u_{\lf \vr_c,t \rg}^{\vq_c}
\rangle_{cell} \nn \\  &= - \p_t   {\rm arg} \lq \Phi\lf \vq_c, t
\rg \rq  + {\cal T}\lf \vr_c, \vq_c, t \rg + \dot{{\vr_c}}  \cdot
\overrightarrow{{\cal R}}\lf \vr_c, \vq_c, t \rg , \label{KinEn}
\eeqa where we have introduced the connection components ${\cal
T}$ and $\overrightarrow{{\cal R}}$ in analogy with $\vec \cQ$ in
(\ref{BerryConn}).  The first term in (\ref{KinEn}) involving the
partial time derivative of the phase in the quasi-momentum
distribution can be rearranged in terms of the total time
derivative \beqa \p_t {\rm arg} \lq \Phi\lf \vq_c, t \rg \rq  =
\frac{d}{d t }{\rm arg} \lq \Phi\lf \vq_c, t \rg \rq -
\dot{\vq}_{\; c} \cdot
\nabla_{\vq_{\, c}} {\rm arg} \lq \Phi\lf \vq_c, t \rg \rq \nn \\
=
 \frac{d}{d t }{\rm arg} \lq \Phi\lf \vq_c, t \rg \rq +
\dot{\vq}_{\; c} \cdot \lq \vr_c - {\vec{\cA}}\,{\lf \vr_c, \vq_c
, t \rg }\rq  , \eeqa where eq. (\ref{ApproxPos}) has been used.
Thus, taking account  the third term  in (\ref{KinEn}), the
final expression of the approximate Lagrangian will contain only linearly
first  derivatives  of the
 ``generalized coordinates" $\lf \vr_c, \vq_{\, c} \rg $. In fact,
 dropping total time derivatives,
allows us to write down a Lagrangian for a point - like classical
 particle
\beqa L_{app} =
 &  \dot{\vr_c} \cdot \lf \vq_{\, c} +
 \overrightarrow{{\cal R}}\lf \vr_c, \vq_c, t \rg\rg  +
  \dot{\vq}_{\; c}
 \cdot{\vec{\cA}}\,{\lf \vr_c, \vq_c, t \rg } +
 {\cal T}\lf \vr_c, \vq_c, t
 \rg \nn \\ & - {\cE}\lf \vr_c, \vq_c, t \rg  -
 \Delta{\cE}\lf \vr_c, \vq_c, t
 \rg, \label{Lagrapp}
 \eeqa where \beq {\cE} =
 \langle \widetilde{\Psi} | {\hat H}_{\lf \vr_c,t \rg}
  | \widetilde{\Psi}\rangle \quad \textrm{and}\quad \Delta{\cE} =
   \half \langle \widetilde{\Psi} |  \lq  \p_f {\hat H}\,
\nabla_{\vr_c } f \lf {\vr_c}, t \rg \cdot \lf {\hat \vr} - \vr_c
\rg + h.c. \rq  | \widetilde{\Psi}\rangle . \eeq  In general, the
last expression  is quite involved \cite{Niu}, but it is easy to
check its elegant form when applying an external electromagnetic
field represented by the potentials $\lf \vA\lf \vr, t \rg,
V_{el}\lf \vr, t \rg \rg $. In fact, for slowly changing vector
potentials, the solution of the approximate Bloch eigenvalue
problem  \beq \lq \frac{1}{2 m} \lf \hat{\vp} + e\, \vA\lf \vr_c,
t\rg \rg^2 + V_{latt}\lf \vr, t\rg -  e \, V_{el}\lf \vr_c, t \rg
\rq \psi = E \psi \eeq can be written in terms of the function
\beqa {\psi'\, }_{\lf \vr_c,t \rg}^{\vq_c} \approx  \exp\lq i \lf
\vq - e \vA\lf \vr_c, t \rg \rg\cdot\vr \rq  u_{\lf \vr_c,t
\rg}^{\vq_c},\eeqa where $u_{\lf \vr_c,t \rg}^{\vq_c}$ is the
periodic part of the Bloch solution for the same crystal in the
absence of a magnetic field. Because of its definition in
(\ref{BerryConn}),  a space - time dependent change of phase does
not have any influence on the Berry connection. Moreover, $
{\vec{\cA}}\,{\lf \vr_c, \vq_c, t \rg } ={\vec{\cA}}\,\lf \vk_c
\rg, $ where
 \beq\vk_c = \vq_c + e\,\vA\lf \vr_c, t \rg
 \eeq
is the gauge
invariant quasi-momentum. On the other hand, from (\ref{KinEn})
one has $\overrightarrow{{\cal R}}\lf \vr_c, \vk_c, t \rg \simeq
\, e\, \nabla_{\vr_c}\lf{{\vA}}\,\lf \vr_c, t \rg   \cdot \vr_c
\rg - e\, {{\vA}}\,\lf \vr_c, t \rg $ and $ {\cal T}\lf \vr_c,
\vk_c, t \rg \simeq    e\, \p_t {{\vA}}\,{\lf \vr_c, t \rg \cdot
\vr_c }$. Furthermore, the expressions for the semiclassical
approximate energy band and its first order correction take the
form
 \beq {\cE} = {\cE}_0\lf \vk_c
\rg +  e \,V_{el}\lf \vr_c, t \rg \qquad \; \Delta{\cE} =  - {\vec
M}\lf \vr_c, \vk_c, t \rg \cdot \vB \lf \vr_c, t \rg ,\label{EMint} \eeq  where
${\vec M}\lf \vr_c, \vk_c, t \rg = \frac{- e}{2 m_e} \langle
\widetilde{\Psi} | {\hat {\vec L}}
  | \widetilde{\Psi}\rangle  $  is the
local mean magnetic moment of the wave-packet and $\vB
\lf \vr_c, t \rg = \nabla_{\vr_c} \times \vA\lf \vr_c, t \rg$ is
the usual expression of the mean magnetic  field acting on the
wave - packet. This is subject also the mean electric field $\vE
\lf \vr  , t \rg = \p_t {\vec A} \lf \vr  , t \rg - \nabla_{\vr  }
V_{el}\lf \vr  , t \rg$.

Finally, let us observe that the discrete symmetry properties of
the crystals induce restrictions in the expressions of the above
Berry's connections.  In particular, the time reversal invariance
implies the transformation $ \vk \rightarrow - \vk$ and  ${\vec
\Theta}\lf \vr , -\vk\rg = - {\vec \Theta}\lf \vk \rg $. So always
${\vec \Theta}$ vanishes at $\vk ={\vec 0}$. Moreover,  the
spatial inversion  implies that ${\vec \Theta}\lf -\vk\rg = {\vec
\Theta}\lf \vk \rg $.  The simultaneous space - time inversion
symmetry implies  ${\vec \Theta}\lf -\vk\rg \equiv 0$. In
conclusion, crystals admitting time and space inversions cannot
carry any dual magnetic structure. On the contrary,  there exist
concrete examples of crystals  for which one, or both, inversion
invariances are broken, so the geometric phase effects are
significant \cite{Niu,Fang}.

%%%%%%%%%%%%%%%%%%%%%%%%%%%%%%%%%%%%%
\section{Hamiltonian Structure}
%%%%%%%%%%%%%%%%%%%%%%%%%%%%%%%%%%%%%%%%
From the semiclassical Lagrangian (\ref{Lagrapp}) we can derive
the equations of  motion of the wave-packet
 \beqa \lf 1 + \Xi \rg\, \dot{\vr}  +
\Theta\,  {\dot{\vq  }}  & = &
 \nabla_{\vq  } \lq \cE
+ \Delta{\cE} - \cT \rq + \p_t {\vec \cQ}, \nn \\  X \, {\dot{\vr}
} + \lf 1 + \Xi  \rg\,\dot{\vq}   & = & -\nabla_{\vr  } \lq \cE +
\Delta{\cE} - \cT \rq  -\p_t {\vec \cR} ,\label{eqmotgen}\eeqa
(dropping the index ``$c$" for simplicity) where the antisymmetric
matrices $\Xi$ and $X$ have  elements  \beqa \Xi_{i, j}\lf \vr ,
\vq , t \rg = \p_{r_i} \cQ_j - \p_{ q_j} \cR_i = i \lf \langle
\nabla_{\vq }\, u_{\lf \vr ,t \rg}^{\vq } | \times | \nabla_{\vr }
\, u_{\lf \vr ,t \rg}^{\vq } \rangle_{cell}\rg_{i j}, \nn
\\    X_{i, j} \lf \vr  , \vq  , t \rg =
 \p_{r_i}\cR_j - \p_{r_j}\cR_i = i \lf \langle \nabla_{\vr  }\, u_{\lf \vr  ,t \rg}^{\vq  } | \times
| \nabla_{\vr  } \, u_{\lf \vr  ,t \rg}^{\vq  } \rangle_{cell}
\rg_{i j}. \eeqa We   also have  \beq \p_t {\vec \cQ} -
\nabla_{\vq } \cT = 2 {\rm Im} \langle \nabla_{\vq  } u_{\lf \vr
,t \rg}^{\vq } |
 \p_t  u_{\lf \vr  ,t \rg}^{\vq  }  \rangle_{cell},
 \qquad \nabla_{\vr  } \cT  -\p_t {\vec \cR} = - 2 {\rm Im} \langle
  \nabla_{\vr  } u_{\lf \vr  ,t \rg}^{\vq  } | \p_t  u_{\lf \vr  ,t \rg}^{\vq  }
   \rangle_{cell}  . \label{altriTens}\eeq
This dynamical system is defined on the tangent manifold $T M$ of
the configuration space, parametrized by the generalized coordinates
$\overrightarrow{\xi} = \lf \vr  , \vq  \rg$.

The system (\ref{eqmotgen}) can be written (at least   when $ \p_t {\vec \cQ} =
\p_t {\vec \cR} = 0 $)
in terms of the symplectic 2-form \beq
\omega = \lf \delta_{i,j}+ \Xi_{i j}\,\rg d r_{\;i} \wedge d
q_{\;j} + \frac{1}{2} \lq X_{i j} \, d q_{\;i} \wedge d q_{\;j} -
\Theta_{i j} \, d r_{\;i} \wedge d r_{\;j} \rq \label{TQHam}\eeq
and  the Hamiltonian function
\beq
{\cal H} = \cE + \Delta{\cE} -
\cT
\eeq
 in the form $i_{\Delta} \omega = d {\cal H}$
where  $
\Delta = {\dot{\vr}  } \frac{\p}{\p {\vr  }} + {\dot{\vq}  }
\frac{\p}{\p {\vq  }}$ \cite{SSD, Marmo}.
The motions   can be viewed therefore,
 as the integral curves of  the vector field
 $\Delta$.
%(Let us note that when the quasi-momentum
%variables $q_i$ run on a torus $T^d$,
% the Hamiltonian vector-fields leading to equations
%  (\ref{eqmotgen}) (or (\ref{TQHam})
%are only locally Hamiltonian.)

 Furthermore,  we assume that the
$2$-form $\omega$   closed, (i.e.  $d \omega = 0$).
% on $T M$.
The closure condition on $\omega$
 is equivalent to the set of differential constraints  \beqa
 \varepsilon_{i,j,k}\,  \p_{q_i} \Theta_{j,k} = 0, \qquad &
 \varepsilon_{i,j,k}\,  \p_{r_i} X_{j,k} = 0 ,  \nn \\
 \p_{q_j} \, \Xi_{i,j} = - \, \p_{r_j} \, \Theta_{i,j} ,
 \qquad & \p_{r_j} \, \Xi_{i,j} =  \, \p_{q_j} \, X_{i,j},
 \\ \lf 1-\delta_{h, k}\rg \varepsilon_{k, i, j}\, \p_{q_k}
 \Xi_{i,j} =  \varepsilon_{h, i, j}\, \p_{r_h}
 \Theta_{i,j},\quad & \lf 1-\delta_{h, k}\rg \varepsilon_{k, i, j}\, \p_{r_k}
 \Xi_{i,j} = - \varepsilon_{h, i, j}\, \p_{q_h}
 X_{i,j}\nn
 .\label{Closure}\eeqa Because of the antisymmetry and the
differentiability properties of the  tensors $\Theta$, $\Xi$ and
$X$ as defined in (\ref{BerryTens}) and (\ref{altriTens}),
 the equations above are automatically satisfied
 for a variational system. In particular,
let us observe that the first pair of equations in (28) are the
divergenceless conditions for the two vector-fields
$\varepsilon_{i,j,k}\, \Theta_{j,k} $ and $\varepsilon_{i,j,k}\,
\p_{r_i} X_{j,k} $ in the $\vq$- and $\vr$-space, respectively.
This  will take a precise physical meaning in the case of
Bloch electrons interacting with external electromagnetic fields.

If $\omega$ is non degenerate, it can be inverted and the
system takes a Hamiltonian form  \cite{SSD, Marmo}.
A non-degenerate and closed 2-form $\omega = \omega_{\alpha,
\beta}\,  d \xi_{\alpha} \wedge d \xi_{\beta}$ defines indeed
a Poisson bracket. For
any pair of functions $f\lf \vr, \vq \rg$ and
$g\lf \vr, \vq \rg$ is associated
$
\lgr f, g \rgr =
\omega^{\alpha, \beta} \p_{\alpha} f \p_{\beta} g,
$
 where
$\omega^{\alpha, \gamma} \omega_{\gamma, \beta} =
\delta^{\alpha}_{\beta}$ is the inverse of the symplectic matrix.
Then Hamilton's equations read
\beq
\dot{\xi}_\alpha=\{\xi_\alpha,{\cal H}\}.
\label{Hameq}
\eeq
In the degenerate case one has to resort to
 symplectic reduction  \cite{SSD,FaJa}.

In the present case here one has a $6\times 6$  block matrix
\beqa \lf \omega_{\alpha, \beta} \rg =\frac{1}{2} \left(%
\begin{array}{cc}
  X  & {\bf 1} + \Xi \\
-{\bf 1} + \Xi & - \Theta \\
\end{array}
\right)
\eeqa
which  is non degenerate when $1-\frac{1}{2}{\rm Tr}
 \lf \Xi ^2 + X \lf {\bf 1} +
2 \, \Xi \rg \Theta \rg \neq 0$.
Then the inverse of the symplectic matrix is
\beqa
&\lf \omega^{\alpha, \beta} \rg  =
-2\Big(1 -  \frac{1}{2}  {\rm Tr} \lf \Xi ^2 + X \lf {\bf 1} +
2 \, \Xi \rg \Theta \rg\Big)^{-1}
\\[12pt]
%\frac{
&\lf {%\tiny
\displaystyle\begin{array}{cc} \Theta +\lq \Theta, \Xi \rq & \lq 1 -
\frac{1}{2}
 {\rm Tr } \lf \Xi^2 + X \, \Theta \rg \rq {\bf 1} + \lf \Xi^2 + X \,
 \Theta \rg^T
 \\[18pt]
-\lq 1 - \frac{1}{2} {\rm Tr }
\lf  \Xi^2 + X \, \Theta \rg \rq {\bf 1} -  \lf  \Xi^2 + X \, \Theta \rg  & - X + \lq \Xi, X \rq \\
\end{array}}
\right).
%}{1 -  \frac{1}{2}  {\rm Tr} \lf \Xi^2 + X \lf {\bf 1}+2\, \Xi \rg \Theta \rg.
\nn
\label{Poisson}
\eeqa

Consistently with the Darboux's theorem on non degenerate
symplectic forms, canonical coordinates i. e. such that
$\omega_{\alpha\beta}=\left(\begin{array}{cc}0&1\\-1&0\end{array}\right)$
can be found.
Then the equation (\ref{Hameq}) takes the usual canonical form;
the disadvantage is that the
 Hamiltonian may become rather complicated.

It should be noticed that  our description is formulated on the
tangent space $TM$, instead of the cotangent bundle $T^*M$,
 usually used in the theory a Hamiltonian system.

 A general  consequence
of the geometrical  formulation is that the Liouville theorem remains true for the
Bloch electron wave-packets and the  invariant  volume is modified with respect
 to   standard semiclassical approximation form
\cite{Ashcroft} as follows  \begin{equation}
 \sqrt{\det\lf \omega_{\alpha\beta}\rg}\,\prod dr_i\wedge dq_i = \frac{1}{8}
 \lf 1 -  \frac{1}{2}  {\rm Tr} \lf \Xi ^2 + X \lf {\bf 1} +
2 \, \Xi \rg \Theta \rg \rg \prod_{i}
 dr_i\wedge dq_i.
 \label{exovol}
 \end{equation}

Before proceeding, let us notice that, in  presence of an external
electromagnetic field, eq.s (\ref{eqmotgen}) take the form  \beqa
\dot{\vr} & = & \nabla_{\vq  } \lq \cE_0\lf \vk \rg - {\vec M}\lf
\vr  , \vk , t \rg \cdot {\vec B}\lf \vr , t \rg\rq -
{\dot{\vk}  }\times {\vec \Theta}\lf  \vk   \rg, \nn \\
\dot{\vk}   & = & - e \lf {\dot{\vr}  }\times {\vec B}\lf \vr  , t
\rg + {\vec E}\lf \vr  , t \rg\rg + \nabla_{{\vr}  }\lf {\vec
M}\lf \vr  , \vk  , t \rg \cdot {\vec B}\lf \vr  , t \rg \rg  ,
\label{EqMot} \eeqa where  the so-called ``dual'', or ``reciprocal
magnetic field'' $ { \Theta}_i\lf \vk  \rg = \half \epsilon_{i j
k} \Theta_{j, k}\lf \vk \rg$ has been introduced.

The  electron mean velocity ${\vec v}= \langle \Psi|
\nabla_{\vk}{\hat{H}}|\Psi \rangle \approx  \langle\tilde{ \Psi}|
\nabla_{\vk}{\hat{H}}|\tilde{\Psi} \rangle $ can be estimated from
the eq.s (\ref{EqMot}). In particular, this can explain the
Anomalous Hall Effect predicted by Karplus and
 Luttinger half a century ago \cite{AHE}. In fact,
 for vanishing
external magnetic field  and choosing the electric field $\vE =
E  {{\hat x}}$, one obtains  the  \beq \left\{%
\begin{array}{l}
    \dot{\vr} =  {\vec v} \lf \vk \rg = \nabla_{\vk} \cE_0\lf \vk \rg +
    e\,
    E \, \Theta_z\lf \vk \rg {{\hat y}}, \\
   {\vk}  =  \vk_{0} - e \,  E  \,t\, {{\hat x}} . \label{anomalHall}\\
\end{array}%
\right.    \eeq  At the macroscopic level, for static uniform
electric and  temperature fields, the electric current will be
given by the mean value   \beq {\vec j} = -e \, \int_{IBZ}
\frac{d^3 \vk}{4 \pi^3} {\vec v}\lf  \vk \rg  g\lf \vr, \vk \rg
\label{current}\eeq with respect to the appropriate  non
equilibrium distribution function $g\lf \vr, \vk \rg$. A good
approximation \cite{Ashcroft} of such a distribution  is  $ g\lf
\vr, \vk \rg = f_{FD}\lf  \cE \rg +  e\, \tau\lf \cE
 \rg  \frac{\p f_{FD}}{\p \cE} \vE \cdot {\vec v}
$, where $f_{FD}\lf \cE \lf\vk\rg \rg$ denotes  the Fermi - Dirac
distribution and   $\tau$ is the relaxation time, that is the
inverse of the  probability for unit time that an electron of
momentum $\vk$ be  scattered. In the simplest situation of  a
filled band, i.e. for $f_{FD} \equiv 1$, the current expression
(\ref{current}) will contain only  the Hall contribution, which
takes the form  \beq j_y = - e^2 E \int_{IBZ} \frac{d^3 \vk}{\lf 2
\pi \rg^3} \Theta_z\lf \vk \rg .\label{AnomalHallcurr}\eeq Thus,
this ``anomalous" Hall effect is entirely due to the existence of
the reciprocal magnetic field ${\vec \Theta}$. Moreover, for a
prismatic fundamental cell in the reciprocal lattice, the
expression (\ref{AnomalHallcurr}) is proportional to the
$k_z$-integrated flux of $\Theta_z$ through the surface,  obtained
by the intersection of the $IBZ$ with the $\lf k_x , k_y \rg$
plane. In fact, from the curvature nature of the integrand (see
eq. (\ref{BerryTens})), such a  flux is interpreted as the first
Chern number (that is a topological invariant taking integer
values) of the principal bundle generated by attaching to each
point of the $IBZ$ an  $U\lf 1 \rg$-fiber, to which the Berry
connection belongs \cite{Thou,Avr}.

%%%%%%%%%%%%%%%%%%%%%%%%%%%%%%%%%%%%%%%%%
\section{Symmetry Group and Symplectic Structure }
%%%%%%%%%%%%%%%%%%%%%%%%%%%%%%%%%%%%%%
Now,  equations of motion of the type
(\ref{EqMot})  in (2+1)-dimensions  were studied
 by several authors from the point of
view of symmetries, in particular, starting from the planar
Galilei group \cite{exotic}.  This  admits a two-fold  central
extension, labeled by the mass $m$  and the ``exotic'' parameter
$\kappa$. The latter can be viewed as a particular case of
Berry's phase, as it is involved in the model considered in
\cite{DH} and
 \cite{HMS}, where the coupling to  an external
electromagnetic field was also  considered.

In the 2-dim gauge-invariant momentum space we consider  the
linear Berry's connection  ${\vec{\cA}}_i\,\lf \vk \rg = -
\frac{\theta}{2}\,\varepsilon_{i, j} k_j $,  modulo further gauge
transformations in $\vk$. This leads to the
 constant Berry's curvature ${ \Theta_3} = - \, \theta $, which
 will be
related to the ``exotic'' charge $\kappa$ of the symmetry Galilei
group.  Then, the first-order Lagrangian (\ref{Lagrapp})  is
simplified to
\begin{equation}
     L=
     \vk \cdot\dot{\vr}-\frac{\vk^2}{2m}
     +e(\vA \cdot \dot{\vr}+V_{el})
     + \frac{\theta}{2}\, \vk \times \dot{\vk},
     \label{totlag}
\end{equation}
where the electromagnetic field depends only on $\lf r_1,r_2, t
\rg$.   The corresponding Euler-Lagrange equations, specializing
(\ref{EqMot}) , are \beq
     m^* \dot{\vr} = \vk - e\, m \,\theta \,{\hat \epsilon }\,
     \vE,\qquad
     \dot{\vk}   =  e B \, {\hat \epsilon }  \, \dot{\vr} + e \, \vE ,
\label{vitesse} \eeq  where the effective mass $ m^*=m(1-e\theta
B) $ appears and ${\hat \epsilon }$  denotes the operator which
rotates vectors of the plane  counterclockwise by $\pi/2$. In the
Hamiltonian framework the modified Poisson-brackets (29) and (32)
become
\begin{equation}
     \begin{array}{ccc}
    \{r_{i},r_{j}\}=\displaystyle\frac{m}{m^*}\,\theta \epsilon_{ij},
    &\{r_{i},k_{j}\}=\displaystyle\frac{m}{m^*}\,\delta_{ij},
    &\{k_{i},k_{j}\}=\displaystyle\frac{m}{m^*}\, e B \epsilon_{ij}.
     \end{array}
     \label{Bcommrel}
\end{equation}
The first important remark is that the Jacobi identity for
(\ref{Bcommrel}) is identically satisfied for arbitrary space -
time dependent magnetic fields. This can be directly checked
restricting  the equations (\ref{Closure}) in 2 dimensions.

The second  feature of the model is that when $m^*=0$, i. e. when
the magnetic field takes the critical value
\begin{equation}
     B=B'_{crit}=\frac{1}{e\theta},
     \label{critB}
\end{equation}
the system becomes singular, and the only allowed motions  follow
the Hall law  \cite{DH}. This is an example of degeneracy of the
symplectic 2-form $\omega$ in (\ref{Poisson}).
   Application of  a uniform and constant  magnetic field
 $B = B'_{crit}$ amounts to
restrict the motion  to the lowest Landau level, and quantization
allows  to recover the ``Laughlin'' wave functions \cite{DH}.
Furthermore, the vanishing of $m^*$ signals a sort of ``phase
transition'',  in the sense that the spectrum of the angular
momentum can take only strictly positive integer values for $m^*
\leq 0$ \cite{HorPlu}.

Now, aiming to describe the system by a larger symmetry group,
following  \cite{Negro} and \cite{HMS}, let us consider a
homogeneous
 time dependent electric field $\vE\lf t \rg$   and its canonical conjugate
momentum  $\vec{\pi}$ as further dynamical variables.  Thinking to
the components $\pi_{i}$ as Lagrange multipliers,  we define the
new Lagrangian \beq L^{enl}=L + {\vec \pi} \cdot \dot{\vE},
\label{enlargLag}\eeq
 providing the supplementary  equations of motion  \beq
\dot{\vE} = 0, \qquad \dot{\vec \pi} = e \vr ,\label{Electric}
\eeq i.e. the electric field is actually a constant. The
Hamiltonian structure is ``enlarged''  appending
$\{E_{i},\pi_{j}\}=\delta_{ij}$ to the fundamental Poisson
brackets (\ref{Bcommrel}).  The  Hamiltonian is $ H_0 = \vk ^2/2 m
- e \vE \cdot \vr $,  for a suitable choice of gauge for $V_{el}$, and
provides the equations of motion (\ref{vitesse}) and
(\ref{Electric}).

 The enlarged
Lagrangian (\ref{enlargLag}) is (quasi-)invariant w. r. t. the
infinitesimal variations
\begin{equation}
     \begin{array}{lcccc}
    \texttt{translations} & \delta \vr = {\va},
    &\delta \vk = 0,
    &\delta \vE = 0,
    &\delta {\vec{\pi}}= e\, {\va} \, t,\hfill
\\[6pt]
  \texttt{rotations}&  \delta \vr = -\phi\, {\hat \epsilon}\,\vr,
    &\delta \vk = -\phi\,  {\hat \epsilon}\,\vk,
    &\delta \vE = -\phi\, {\hat \epsilon}\,\vE,
    &\delta {\vec \pi} = -\phi\, {\hat \epsilon}\,{\vec \pi},\hfill
\\[6pt]
  \texttt{boosts } & \delta \vr = \vb \, t,
    &\delta \vk = m \vb,
    &\delta \vE = - B {\hat \epsilon}\,\vb,
    &\delta \vpi = \frac{e}{2} \,\vb\, t^2 ,\hfill
\\[6pt]
  \texttt{electric} & \delta \vr = 0,
    &\delta \vk = 0,
    &\delta \vE = \vd,
    &\delta \vpi = 0
    ,\hfill\\\texttt{superposition}& & & & \hfill
    \end{array}
\label{engal}
\end{equation}
where the 2-components vectors $\va,\; \vb, \; \vd $ are  parameters  related to the space translations,
boosts and  linear changes of the electric field, respectively.
The scalar $\phi$ is the  rotational  parameter.

 Conserved quantities are readily constructed by Noether
theorem. Actually,
  direct integration of the second equation in (\ref{vitesse})  yields the
constant of the motion
\begin{equation}
     {\vec{\cal P}} = \vk - e B \he \vr - e \vE\, t,
    \label{enlargedmom}
\end{equation}
describing   uniform motions of the guiding center of the charged
particle. Using the commutation relations
\begin{equation}
     \{r_{i},\cP_{j}\} = \delta_{ij},
     \qquad
     \{k_{i},\cP_{j}\}=0,
     \qquad
     \{E_{j}, \cP_{i}\}=0
     \qquad
     \{\pi_{i},\cP_{j}\} = e t\,\delta_{ij},
\end{equation}
  one recognizes  that (\ref{enlargedmom})
 generates  the enlarged translations in (\ref{engal}).
Similarly, \beq
     \cJ= \vr \, \times \vk +
     \frac{e B}{2}\,{\vr} \,^2 +
     \frac{\theta}{2}\,{\vk} \,^2 + \vE \times {\vec \pi} + s_{0},
     \quad
     {\vec \cK} = m\, \vr - \left({\vec  \cP} + \frac{ e\, \vE \, t}{2}\right)\, t
     + m \, \theta\, \he \, \vk - B \he\;{\vec \pi}.    \label{enlargedboostCC}
\eeq are conserved quantities  and generate rotations and  boosts,
respectively, accordingly with (\ref{engal}). In
(\ref{enlargedboostCC}) the ``anyonic" spin $s_{0}$ has been added
by hand to the angular momentum $\cJ$,  which contains also  the
magnetic flux and new ``exotic'' (or ``dual") flux, proportional to
the ``area'' swept in momentum space. Of course also the electric
field $\vE$ is considered as a further conserved quantity.
%
%%%%%%%%%%%%%%%%%%%%%%%%%%%%%%%%%%%%%%%%%%%%
%\subsection{Enlarged planar Galilei algebra}
%%%%%%%%%%%%%%%%%%%%%%%%%%%%%%%%%%%%%%%%%%%%
%
Together with the Hamiltonian $H_0$, they span a  11-dimensional
symmetry Lie algebra, whose non vanishing Lie-Poisson
commutation relations are
\begin{equation}
\begin{array}{ccc}
   \{\cP_{i},H_0\}= e E_{i},\hfill
   &\{\cK_{i},H_0\}= \cP_{i},\hfill
   &\{\cP_{i},\cJ\}= - \epsilon_{ij}\,\cP_{j},\hfill
    \\[10pt]
    \{\cP_{i},\cP_{j}\}=-e\, B\,\epsilon_{ij},\quad\hfill
    &\{\cK_{i},\cJ\}=-\epsilon_{ij}\,\cK_{j},\hfill
    &\{\cP_{i},\cK_{j}\}=-m\,\delta_{ij},\hfill
    \\[10pt]
   \{\cK_{i},\cK_{j}\}=- \theta\,m^2\,\epsilon_{ij},\quad\hfill
   &\{E_{i},\cJ\} = - \epsilon_{ij}\, E_{j},\quad
   &\{E_{i},\cK_{j}\} = B\,\epsilon_{ij},\quad\hfill
    \end{array}
    \label{enlargedGal}
\end{equation}
  with $m$ and $B$  as  central
  charges.  But the key observation is that the boosts generators
$\cK_i$ do not  commute among them-selves as usual. In fact,
  their commutator  yields the ``exotic'' central charge
$\kappa=-\theta\, m^2$
and, then, providing us with an explicit realization
  of the second central extension of the Galilei group.

The action of the symmetry group  on these functions, formally
belonging to the dual space of the symmetry algebra,  is given by
\beqa & H_0' = H_0 - \vb \cdot R_{\phi} {\vec \cP} + \frac{1}{2} m
\vb^2 - e\,\va \cdot  \vE' ,\qquad
{\vec \cP}' =  R_{\phi} {\vec \cP}  + e \tau \vE'  - e B \, \he \va - m \,\vb ,&\nn \\
\cJ' &= \cJ + \half e B \,\va^2 - m \, \va \times \vb + \half \,
m^2\, \theta \, \vb^2 + \va \times R_{\phi} {\vec \cP} + \vb
\times R_{\phi} {\vec \cK} + \lf R_{\phi} \vd  + e \, \tau \, \va\rg \times \vE' ,& \nn \\
 &{\vec \cK}' =  R_{\phi} {\vec \cK} + m \va +\tau m \vb - m^2 \theta \, \he \vb
 + \half e  \tau^2 \vE' - \tau  R_{\phi} {\vec \cP}  , \qquad
 \vE ' =   R_{\phi} \vE + B \he \, \vb ,&\eeqa where $ R_{\phi}$
 represents the plane rotation by an angle $\phi$
and the  parameter $\tau$ is  the time translation. Besides $m$,
$\kappa = - m^2 \theta $ and $B$, the  enlarged
 Galilei group has the
independent Casimir functions \beqa
     { %\tiny
      \cC\hfill = e\theta\left(B H_0 -   {\vec \cP } \times {\vE}
     + \frac{m}{2 B}\,\vE^2 \right) = \frac{e\theta B}{2m}
     \lf k_{i}- \frac{m}{B}\epsilon_{ij}E_{j}\rg^2,}\hfill  \\
     { \tiny
      \cC'\hfill  = \frac{{\vec{\cP}}^2}{2m} - H_0 - \frac{e}{m}
     \Big({\vec \cK} \cdot \vE + \cJ B\Big) - \frac{me\theta}{2B}\vE^2 =
     -\cC - \frac{es_{0}B}{m},\hfill}
     \label{casimir1}
\eeqa where $\cC'$ is interpreted as the  internal energy of the
system and $\cC + \cC'$ is the spin, expressed in an energy scale.
 These  are non trivial convex Casimir functions, so they restrict
the group orbits in the dual of the symmetry algebra to
6-dimensional manifolds, except at the critical point defined by
$\cC = 0 $, where they
 define a 4-dimensional manifold. All these submanifolds can be endowed
with suitable Poisson structures. The 6-dim  orbits
labeled by the $\lf m, \kappa, B, \cC \lf \neq 0 \rg,
s_0 \rg$ are endowed by local coordinates $\lf {\vec \cP}, {\vec
\cK}, \vE \rg $. The restricted Hamiltonian $H_0$  to such an
orbit becomes linear in the momenta $\cP_i$, i.e. \beq H_1 =
\frac{{\vec \cP} \times \vE}{B} - \frac{m}{ 2 B^2} \vE^2 +
\frac{\cC}{e\, \theta\, B}. \label{Ham1}\eeq The relevant Poisson
brackets  are extracted from (\ref{enlargedGal}), yielding  the  non
singular symplectic form on the orbit \beq \omega_1 =
\frac{1}{B} d\cP_1 \wedge d\cP_2 + \frac{m}{B^2 \,e} d\cP_i \wedge
d E_i - \frac{\epsilon_{i j}}{B}\, d\cK_i \wedge d E_j + \frac{m
\, m^*}{B^3\,e} d E_1 \wedge d E_2. \eeq On such an orbit, the
equations of motion
 can be read off directly from the first to
relation in (\ref{enlargedGal}). Their immediate solution \beq
{\vec \cP} = e \vE_0 t + {\vec \cP_0},
\quad {\vec \cK} = \half \vE_0
t^2 + {\vec \cP_0 }t + {\vec \cK_0 },
\quad
\vE = \vE_0 ,  \eeq
describes the solutions of the original equations
of motion (\ref{vitesse}) in terms of the new variables.   Thus
one obtains the usual cycloidal motions, with guiding center,
 radius and frequency, given in terms of integrals of motion  by
\beq \vr_0 = \frac{1}{B^2\, e}\lf m \, \vE_0 + B \,\he {\vec \cP_0}
\rg + \he \, \frac{\vE_0}{B}\, t, \quad  \rho = \frac{m}{e\, B}
\sqrt{\frac{\cC}{e \, m \,\theta\, B }}, \quad  \Omega =
\frac{e \, B}{m^*}. \eeq
  The
singular 4 dimensional orbits  can be expressed equivalently by
 \beq {\vec \cP} = \frac{m}{B} \, \he \, \vE + \alpha \, \vE
\qquad \lf \alpha \in \,\mathbf{R} \rg \label{singularorbit}. \eeq
 In terms of
the original variables,  from the second equality in (50) one has
constrained both components of the momentum $\vk$. Consequently,
the  equations of motion (\ref{vitesse}) become \beq \dot{\vr} =
\he \, \frac{\vE}{B}, \qquad \vk = \frac{m}{B}\,\he \, \vE
,\label{Hall}\eeq as predicted by (\ref{critB}) at $\; m^* =0 \;
\Leftrightarrow \; B = B'_{crit}$.
 Notice also that   the above
formulae give  vanishing cycloid radius and diverging  frequency
$\Omega$. In other words, all motions reduce to uniform
translations, driven at the Hall velocity. On the orbits $\lf m,
\kappa, B,\cC = 0 , s_0 \rg$ naturally the  set of coordinates
$\lf {\vec \cK}, \vE \rg$  is introduced, and defined the non
degenerate symplectic 2-form \beq \omega_2 = \epsilon_{i j}\,
d\cK_i \wedge d E_j +   e m^2 \theta^2 d E_1 \wedge d E_2.\eeq The
Hamiltonian becomes $ H_2 = \frac{m}{2 B_{crit}^2} \vE ^2 $ and
yields the equations $ { \dot \vE} = 0, \; { \dot {\vec \cK}} =
\frac{m}{B} \, \he \vE$,  showing again  that the particle motion is reduced
to the uniform translations of the  guiding center.

%%%%%%%%%%%%%%%%%%%%%%%%%%%%%%%%%%%%%%%%%%%%%%%
%\section{Algebraic construction of the coupled anyon plus
%electromagnetic field system and Hall effects}\label{Bacry}
%%%%%%%%%%%%%%%%%%%%%%%%%%%%%%%%%%%%%%%%%%%%%%%

Following a procedure introduced by Bacry \cite{Bacry}, a direct
generalization of the  system (\ref{vitesse})-(\ref{Electric}),
endowed with the enlarged Galilei symmetry, can be derived from
the unique polynomial Hamiltonian
\begin{equation}
H_{anom}=H_0 + \frac{g}{2}\,\cC'=
\frac{\vk{}^2}{2m}\big(1-\frac{g}{2}e\theta B\big)-e\vE\cdot{\vr}
-\mu\,B + \frac{g e \theta}{2}\,\vk \times \vE-\frac{m g e
\theta}{4 B}\vE{}^2 .
     \label{anom'Ham}
\end{equation}
Here $\mu=g \,e\,s_{0}/2m$ with $g$  a real parameter, interpreted
as the anomalous gyromagnetic factor. The kinetic energy
term gets a field-dependent factor, which can be seen  also
 as wave-packet
magnetic dipole interaction, with $M = \frac{ g\, e \,\theta}{4 m} \vk\,^2$, accordingly to (\ref{EMint}) and (\ref{EqMot}). The Hamiltonian
(\ref{anom'Ham}) contains, together with the standard magnetic
moment term $\mu B$, also contributions similar to the Hamiltonian
(\ref{Ham1}), proportional to $g$. However, we have to consider
that we are using now ``natural" coordinates.  The corresponding
equations of motion are
\begin{equation}
m^*\dot{\vr} = \big(1-\frac{g}{2}e\theta B \big)\vk -
     \Big(1-\frac{g}{2}\Big) e m \theta \he \vE , \qquad
     \dot{\vk}   =  e B \, {\hat \epsilon }  \, \dot{\vr} + e \, \vE.
\label{enlvit}
\end{equation} This
 is a special case of (\ref{EqMot}) and  reminiscent of  Eq. (5.3) in  \cite{AnAn}. In particular,  for $g=2$ and $e\theta B\neq1$ one obtains  $m\dot{\vr} = \vk $,
so that our equations describe an ordinary charged particle in an
electromagnetic field. For $g=2$ and $e\theta B=1$, Eq.
(\ref{enlvit}) is identically satisfied.
%%%%%%%%%%%%%%%%%%%%%%%%%%%%%%%%%%%%%%%%%%%%
%\section{Hall effects}\label{Hallsec}
%%%%%%%%%%%%%%%%%%%%%%%%%%%%%%%%%%%%%%%%%%%%
Of course, since the symmetry structure of this new anomalous
system is the same as for the standard case $g = 0$, the analysis
of the motions follows essentially the same considerations as
above. The only change is that the frequency of the rotational
motion is $
     \Omega=\frac{eB}{2m^*}\big(1-\frac{g}{2}e\theta B\big)
  $,
which for $g=2$ reduces to the usual Larmor frequency $eB/m$. But,
now $\Omega$ vanishes at the new critical point $B=B''_{crit} =
\frac{2}{e\, g\, \theta}$. At this value  $m^* = m \lf1 - 2/g \rg$
and the equation (\ref{enlvit}) becomes an identity at $ g = 2$.
On the other hand, for $g \neq 2 $ it reduces to
   $ \dot{\vr}=\frac{g}{2}\,e\theta \he \vE,
$ which  again defines motions following the Hall law
(\ref{Hall}), except that now $\dot{\vk} = 0$, that is  the
momentum is an arbitrary constant.

%%%%%%%%%%%%%%%%%%%%%%%%%%%%%%%%%%%%%%%%%%%%%%%%%%%%%%
% Algebra deformation
%%%%%%%%%%%%%%%%%%%%%%%%%%%%%%%%%%%%%%%%%%

A different kind of generalization can be obtained by a
deformation of the symmetry algebra, although the symmetry
generators remain essentially the same.  In particular, we would
change only the Hamiltonian, combining $H_0$ with the other
generators. Although a systematic study lies beyond the scope of
the present work, let us consider the Hamiltonian with the
magnetic interaction  \beq H_{mag} = H_0 + \mu  B \cJ. \eeq This
model has not to be confused with that introduced at the beginning
in the context of the solid state physics, since $\cJ$ has a
physical different meaning with respect to $\vec M$. However, it
could be useful in some limit.  The Poisson brackets in
(\ref{enlargedGal}) are modified only when commuting
with $H_{mag}$. Specifically,  \beq
\lgr H_{mag}, {\vec \cP}\rgr = - e\, \vE + \mu B\, \he {\vec \cP},
\quad
\lgr H_{mag}, {\vec \cK}\rgr = -  {\vec \cP} + \mu B \he \,
{\vec \cK},
 \quad
 \lgr H_{mag}, \vE \rgr = \mu B \he \, \vE.\eeq
Notice that the presence of the $\vec \pi$  in
$\cJ$  induces the non constancy of the electric field.  Moreover,
one can find only one Casimir operator, namely \beq \cC_{mag} =
\half {\vec \cP}^2 - e B \cJ + \frac{m m^*}{B^2} \vE^2 -e {\vec
\cK} \cdot \vE - \frac{m}{B} {\vec \cP} \times \vE = \frac{m^*}{m}
B^2 \lf \vk -\frac{m}{B} \he\,\vE\rg^2 - 2 e B^2s_0 .\eeq
This invariant is a convex function, and therefore we expect
dimensional reduction of the group orbit as in  (\ref{singularorbit}).
The equations of motion, in term of the  natural coordinates $\vr$ and
$\vk$, take the form \beqa m^* \dot{\vr} &= \vk - e \, m\, \theta
\he \, \vE, \qquad &\dot{\vk} =  e \lf 1 + B \, m\,\theta\,\mu \rg
\, \vE + B \, \lf
e + m^*\, \mu \rg \lf \he\,\dot{\vr} - B \, \mu \vr \rg,\nn \\
\dot{\vE} &= - B\,\mu\, \he \, \vE, \qquad &\dot{\vpi} =
B\,\mu\,\vpi + e\, \vr .\eeqa

%%%%%%%%%%%%%%%%%%%%
\section{Discussion}
%%%%%%%%%%%%%%%%%%%%
In conclusion, the semiclassical Bloch electron provides a physical motivation for the ``exotic `` particle models in the plane, and the ``enlarged Galilean symmetry'' provides us wih
further hints.
 Generalizations
of the ``exotic'' models are built by  adding a Casimir to
the Hamiltonian,  accomodating in certain limits
anomalous moment coupling and orbital magnetic dipole interactions.
An algebraic characterization of the Hall
motions in terms of criticality conditions for the symmetry
Casimir operators was shown.
%Other possible extension related to the
%semiclassical theory of a Bloch electron can be considered in this framework.

%%%%%%%%%%%%%%%%%%%%%%%%%%%%%%%%%
\kikezd{Acknowledgement}  This paper contains a review based on
the  joint researches with C. Duval, Z. Horv{\'a}th
and P. Stichel, to whom we express our indebtedness.
PAH would like to thank Lecce University
for hospitality extended to him.
This work was
partially supported by MURST grant SINTESI-2004 and by the INFN
grant LE41.

%%%%%%%%%%%%%%%%%%%%%%%%%%%%%%%%%%%%%%%%%%%%%%%%%%%%
%%%%%%%%%%%%%%%%%%%%%%%%%%%%%%%%%%%%%%%%%%%%%%%%%%%%%%%%%%%%%%%%%%%%%%%%%%%%%%


\begin{thebibliography}{99}
%%%%%%%%%%%%%%%%%%%%%%%%%%%%%%%%%%%%%%%%%%%%%%%%%%%%%%%%%%%%%%%%%%%%%%%%%%%%%%
%%%%%%%%%%%%%%%%%%%%%%%%%%%%%%%%%%%%%%%%%%%%%%%%%%%%

\bibitem{Ashcroft} N. W. Ashcroft and N. D. Mermin: {\it Solid
State Physics}, Saunders, Philadelphia (1976).

\bibitem{Niu}
M. C. Chang and Q. Niu, {\sl Phys. Rev. Lett}. {\bf 75}, 1348
(1995); {\sl Phys. Rev.} {\bf B 53}, 7010 (1996); G. Sundaram and
Q. Niu, {\sl Phys. Rev.} {\bf B59}, 14915 (1999). See also A.
Bohm, A. Mostafazadeh, H. Koizumi, Q. Niu and J. Zwanziger, {\it
The Geometric Phase in Quantum Systems}. Chapter 12. Springer
Verlag (2003).

\bibitem{AHE}
R. Karplus and J. M. Luttinger, {\sl Phys. Rev}. {\bf 95}, 1154
(1954); T. Jungwirth, Q. Niu, and A. H. MacDonald, {\sl Phys. Rev.
Lett.} {\bf 90}, 207208 (2002); D. Culcer, A. H. MacDonald, and Q.
Niu, {\sl Phys. Rev.} {\bf B 68}, 045327 (2003).

%\bibitem{Novoselov} K. S. Novoselov {\it et al.},
%{\sl Nature Phys. } {\bf 2}, 177 (2006)

\bibitem{Fang} Z. Fang {\it et al.}:
%``The Anomalous Hall Effect and Magnetic Monopoles in Momentum Space",
{\it Science} {\bf
302}, 92 - 95 (2003);
%\bibitem{Onoda}
 S. Onoda {\it et al.},
%`` Intrinsic vs. extrinsic anomalous Hall effect in ferromagnets",
[\texttt{cond-mat/0605580}].

\bibitem{Kramer} P. Kramer, M. Saraceno, {\it Geometry of the Time
Dependent Variational Principle in Quantum Mechanics},  Ed. J.
Ehelers {\it et al.}  Lecture Notes in Physics {\bf 140}
(Springer-Verlag, Berlin (1981)).

\bibitem{exotic}
J.-M.~L\'evy-Leblond, %{\it Galilei group and Galilean invariance}.
in {\it Group Theory and Applications} (Loebl Ed.), {\bf II},
Acad. Press, New York, p. 222 (1972).
% Y.~Brihaye, C.~Gonera, S.~Giller and P.~Kosi\'nski, \texttt{hep-th/9503046} (unpublished);
%D.~R.~Grigore, {\sl Journ. Math. Phys.} {\bf 37}, 240 and {\sl ibid}. {\bf 37}, 460 (1996).

\bibitem{Negro}
J. Negro and M. A. del Olmo,
{\sl Journ. Math. Phys}. {\bf 31}, 2811 (1990);
J. Negro, M.A. Del Olmo, J. Tosiek,
%:`` Anyons, group theory and planar physics",
{\sl  J. Math. Phys.} {\bf  47} (2006).


\bibitem{HMS}
P.~A.~Horv\'athy, L. Martina and P. Stichel
% {\it Enlarged Galilean symmetry of anyons and the Hall effect}.
{\sl  Phys. Lett}. {\bf B 615}, 87  (2005).
 [\texttt{hep-th/0412090}].



\bibitem{LSZ}
J.~Lukierski, P.~C.~Stichel, W.~J.~Zakrzewski,
   {\sl Annals of Physics (N. Y.)} {\bf 260}, 224 (1997)
[\texttt{hep-th/9612017}].

\bibitem{DH}
C.~ Duval and P.~A.~Horv\'athy,
{\sl Phys. Lett.} {\bf B 479}, 284 (2000) [\texttt{hep-th/0002233}];
{\sl J. Phys.} {\bf A 34}, 10097 (2001) [\texttt{hep-th/0106089}].

%\bibitem{DHH}
%C. Duval, Z. Horv\'ath and P. A. Horv\'athy
%% ``Exotic plasma as classical Hall liquid.''
%{\sl Int. Journ. Mod. Phys.} {\bf B 15} No 26,
% 3397-3408 (2001) [\texttt{cond-mat/0101449}].


\bibitem{Laughlin} R. B. Laughlin,
{\sl Phys. Rev.  Lett.} {\bf 50} (1983),
1395.


\bibitem{BM}
A. B\'erard and H. Mohrbach,
%``Monopole and Berry phase in momentum space
%in noncommutative quantum mechanics.''
 {\sl Phys. Rev.} {\bf D 69}, 127701 (2004)
[\texttt{hep-th/0310167}].

\bibitem{HMS2} P. Horv\'athy, L. Martina, P. Stichel,
% ``Comments on spin-orbit interaction of anyons ",
{\it Mod. Phys.Lett. } {\bf A20} (2005), 1177.

\bibitem{Szabo}M. R. Douglas, N.A. Nekrasov, {\it Rev. Mod. Phys}
{\bf 73} (2001) 977;
  R.~J. Szabo,
  %{\it Quantum Field Theory on noncommutative spaces}.
{\it Phys. Rep.} {\bf 378} (2003),  203.

\bibitem{SSD}
J.-M.~Souriau, {\it Structure des syst\`emes dynamiques}, Dunod:
Paris (1970); {\it Structure of Dynamical Systems: a Symplectic
View of Physics}. Birkh\"auser~(1997).

\bibitem{Marmo} R. Abraham, J. Mardsen, {\it Foundations of Mechanics}, Addison-Wesley, Reading (Mass, 1978).
  G. Marmo, E.J. Saletan, A. Simoni, B. Vitale: {\it
Dynamical Systems}, John Wiley \& Sons (Chichester, 1985).


\bibitem{FaJa}
L. D. Faddeev and R. Jackiw,
%{\it Hamiltonian reduction of unconstrained and constrained systems}.
{\sl Phys. Rev. Lett}. {\bf 60} (1988), 1692.

\bibitem{Thou} D. Thouless, M. Kohmoto, M. Nightingale, M. den
Nijs, {\sl Phys. Rev. Lett}. {\bf 49} (1982), 405.

\bibitem{Avr} J.E. Avron, R. Seiler, B. Simon,
%{\it Homotopy and Quantization in Condensed Matter Physics},
{\sl Phys. Rev. Lett}. {\bf 51} (1983), 51.

\bibitem{HorPlu} P.A. Horv\'athy, M.S. Plushchay,
%`` Nonrelativistic anyons in external electromagnetic field",
 {\sl  Nucl. Phys.} {\bf B 714}, 269 (2005)
[\texttt{hep-th/0502040}]

\bibitem{Bacry}
H. Bacry,  {\sl Lett. Math. Phys}. {\bf 1}, 295 (1976).

\bibitem{AnAn}
C.~ Duval and P.~A.~Horv\'athy,
{\sl Phys. Lett}. {\bf B 594} pp. 402-409 (2004),
[\texttt{hep-th/0402191}];
 B. S. Skagerstam and A. Stern,
 {\sl Physica Scripta} {\bf 24}, 493 (1981);
G. Grignani, M. Plyushchay,
and P. Sodano, {\sl Nucl. Phys.} {\bf B464}, 189 (1996)
[\texttt{hep-th/9511072}].



\end{thebibliography}
\end{document}